\documentclass[twocolumn]{aastex701} 
\usepackage[T1]{fontenc}
\usepackage{ae,aecompl}
\usepackage{hyperref}
\usepackage{url}

\usepackage{longtable}
\usepackage{booktabs}
\usepackage{tabu}
\usepackage{graphicx}
\usepackage{multirow}
\usepackage{ulem}
\usepackage{color}
\usepackage{amsmath}

\hypersetup{citecolor=blue, 
            linkcolor=red, 
            menucolor=blue, 
            urlcolor=blue}  

\def \cm{~\rm{cm}}
\def \s{~\rm{s}}
\def \km{~\rm{km}}

\def \K{~\rm{K}}

\def \g{~\rm{g}}
\def \G{~\rm{G}}

\def \erg{~\rm{erg}}

\def \dday{~\rm{day}}

\usepackage{xcolor}
\definecolor{redak}{rgb}{0.9,0.15,0.05}

\shorttitle{Multiple shells in SN 2023ixf}
\shortauthors{Shiran and Soker}

\graphicspath{{./}{figures/}}

\begin{document}

\title{Multiple shells in supernova 2023ixf support the jittering jets explosion mechanism (JJEM)}

\author[0000-0003-0375-8987]{Noam Soker} 
\affiliation{Department of Physics, Technion Israel Institute of Technology, Haifa, 3200003, Israel; soker@physics.technion.ac.il}
\email{soker@physics.technion.ac.il}

\author[0009-0008-3236-1210]{Kobi Shiran} 
\affiliation{Department of Physics, Technion Israel Institute of Technology, Haifa, 3200003, Israel; soker@physics.technion.ac.il}
\email{kobishiran@campus.technion.ac.il}

\date{\today}

\begin{abstract}
Examining the photospheric time evolution of the core-collapse supernova (CCSN) SN 2023ixf from the literature, we identify two (possibly three) evolutionary time periods with constant expansion velocities, which we attribute to two (or three) ejecta shells.  
We find that several CCSN remnants have morphologies with two or more complete or partial shells, compatible with the presence of two (or three) photospheric shells during the photospheric phase of the explosion. Studies have attributed these CCSN remnants to the jittering-jet explosion mechanism (JJEM), which involves two or three energetic pairs of jets participating in the explosion. We, therefore, conclude that the structure of the photospheric shells of SN 2023ixf supports its explosion by the JJEM. This study adds to the accumulating evidence that the JJEM is the primary explosion mechanism of CCSNE. 
\end{abstract}




\section{Introduction} 
\label{sec:intro}

A major question about core-collapse supernovae (CCSNe) is the mechanism of their explosion. There are two intensively studied competing theoretical explosion mechanisms of CCSNe, the delayed neutrino mechanism and the jittering jets explosion mechanism (JJEM).

The research on the delayed neutrino explosion mechanism
focuses on simulating the revival of the stalled shock around the newly-born neutron star (NS) at $r \simeq 150 \km$ by neutrino heating, finding the conditions for an explosion rather than a `failed supernova', and comparing simulated outcomes with observations 
(e.g., \citealt{Bambaetal2025CasA, Bocciolietal2025, BoccioliRoberti2025, EggenbergerAndersenetal2025, FangQetal2025, Huangetal2025, Imashevaetal2025, Laplaceetal2025, Maltsevetal2025, Maunderetal2025, Morietal2025, Mulleretal2025, Nakamuraetal2025, SykesMuller2025, Orlandoetal20251987A, ParadisoCoughlin2025, PowellMuller2025, Tsunaetal2025, Vinketal2025, WangBurrows2025, Willcoxetal2025, Mukazhanov2025, Raffeltetal2025, Vartanyanetal2025}; \citealt{Janka2025}  for a recent review). 
The magnetorotational explosion mechanism operates in very rare cases of rapid pre-collapse core rotation, leading to the launch of one pair of jets along a fixed axis (e.g., \citealt{Shibataetal2025}). At best, it can account for only a small fraction of CCSNe, and studies of the magnetorotational explosion mechanism attribute most CCSNe to the neutrino-driven mechanism; therefore, we group the magnetorotational mechanism with the neutrino-driven mechanism.\footnote{See  \citealt{Janka2025Padova} for a recent talk on the neutrino-driven mechanism: \url{https://www.memsait.it/videomemorie/volume-2-2025/VIDEOMEM_2_2025.46.mp4}; also \url{https://www.youtube.com/watch?v=nRfDPPSmnzI&t=100s}}  

The research of the JJEM has focused since 2024 on finding supporting evidence for jets in CCSN remnants (CCSNRs; \citealt{Bearetal2025Puppis, BearSoker2025, Shishkinetal2025S147, Soker2025G0901, Soker2025N132D, Soker2025RCW89, Soker2025Dust, SokerAkashi2025, SokerShishkin2025Vela, SokerShishkin2025W49B}, for papers since 2025)\footnote{See \citealt{Soker2025Padova} For a talk on the JJEM: \url{https://www.memsait.it/videomemorie/volume-2-2025/VIDEOMEM_2_2025.47.mp4}}. 
The morphologies of many CCSNRs, in particular point-symmetric CCSNRs (those with two or more symmetry axes), strongly support the JJEM (see the summary of results  and reviews  in  \citealt{Soker2024UnivReview, Soker2025Learning}). 
Other studies of the JJEM since 2025 are three-dimensional simulations of jet shaping \citep{Braudoetal2025} and pre-collapse angular momentum fluctuations in the inner core \citep{WangShishkinSoker2025}. 

The explosion process of CCSNe provides only a few observables to distinguish between the two mechanisms.  
A minority of CCSNe that have explosion energies of $E_{\rm ex} \gtrsim 2 \times 10^{51} \erg$ support the JJEM because the neutrino-driven mechanism struggles to reach these explosion energies (for recent reviews see \citealt{Soker2024UnivReview, Soker2025Learning}). This is the case, for example, in many superluminous stripped-envelope supernovae (e.g., \citealt{Kumar2025}). However, the two explosion mechanisms have similar predictions to most other observables of the explosion process itself (before the remnant is spatially resolved; \citealt{Soker2024UnivReview, Soker2025Learning}). In this study, we find that SN 2023ixf has a property that is more compatible with the JJEM.

We use the photosphere radius evolution $R_{\rm ph}(t)$ of SN 2023ixf that \cite{Zimmermanetal2024} calculated (Section \ref{sec:Photosphere}) to claim that jets exploded SN 2023ixf. In Section \ref{sec:Remnants}, we present CCSNRs that have jet-shaped morphologies that are compatible with two or more photospheric shells (depending on the line of sight) as we deduced in Section \ref{sec:Photosphere}. In this study, shells are not necessarily spherical; they can be caps of protrusions termed ears or partial shells, such as the dense boundary of a jet-inflated bubble. We summarize this short study in Section \ref{sec:Summary}. 

\section{Identifying photospheric shells in SN 2023ixf}
\label{sec:Photosphere}

SN 2023ixf is a well-studied nearby CCSN (e.g., \citealt{Hiramatsuetal2023, Soraisametal2023, Bostroemetal2024, Bostroemetal2025, VanDyketal2024B, DerKacyetal2025, JacobsonGalan2025, JacobsonGalanetal2025, Medleretal2025, Parketal2025, Vinkoetal2025, Zhengetal2025}). It attracted attention from the beginning because of its proximity and dense, compact circumstellar material (e.g., \citealt{Bergeretal2023,  Bostroemetal2023, Grefenstetteetal2023, JacobsonGalanetal2023, SmithNetal2023, Tejaetal2023, Laplaceetal2025B}). 
There are uncertainties regarding the progenitor and explosion energy. \cite{Singhetal2024} model the lightcurve with hydrodynamical calculations and estimated a progenitor radius of $R_\ast \simeq 470 R_\odot$ and an explosion energy of $E_{\rm exp} \simeq 2\times 10^{51} \erg$.  
However, they commented that these values are not unique due to modeling degeneracies.
\cite{VanDyketal2024} estimated the progenitor stellar radius to be in the range of $\simeq 921 - 2012 R_\odot$, with a median of $R_\ast = 1389 R_\odot=9.7 \times 10^{13} \cm$. \cite{Hsuetal2025} estimated the progenitor radius to be $R_\ast \gtrsim 950 R_\odot$. \cite{MoriyaSingh2024} estimated the explosion energy to be $E_{\rm exp} \simeq (2-3) \times 10^{51} \erg$, while \cite{Micheletal2025} estimated a more typical energy of $E_{\rm exp} \simeq (0.3-1.4) \times 10^{51} \erg$.

The parameters we take to our study are a typical CCSN explosion energy, and a progenitor radius in the range of $R_\ast \simeq 500 R_\odot - 1400 R_\odot$, or 
$R_\ast \simeq  4 \times 10^{13} - 10^{14} \cm$. 

\begin{figure*}
\begin{center}
\includegraphics[trim=0.0cm 0.50cm 0.0cm 2.0cm ,clip, scale=0.53]{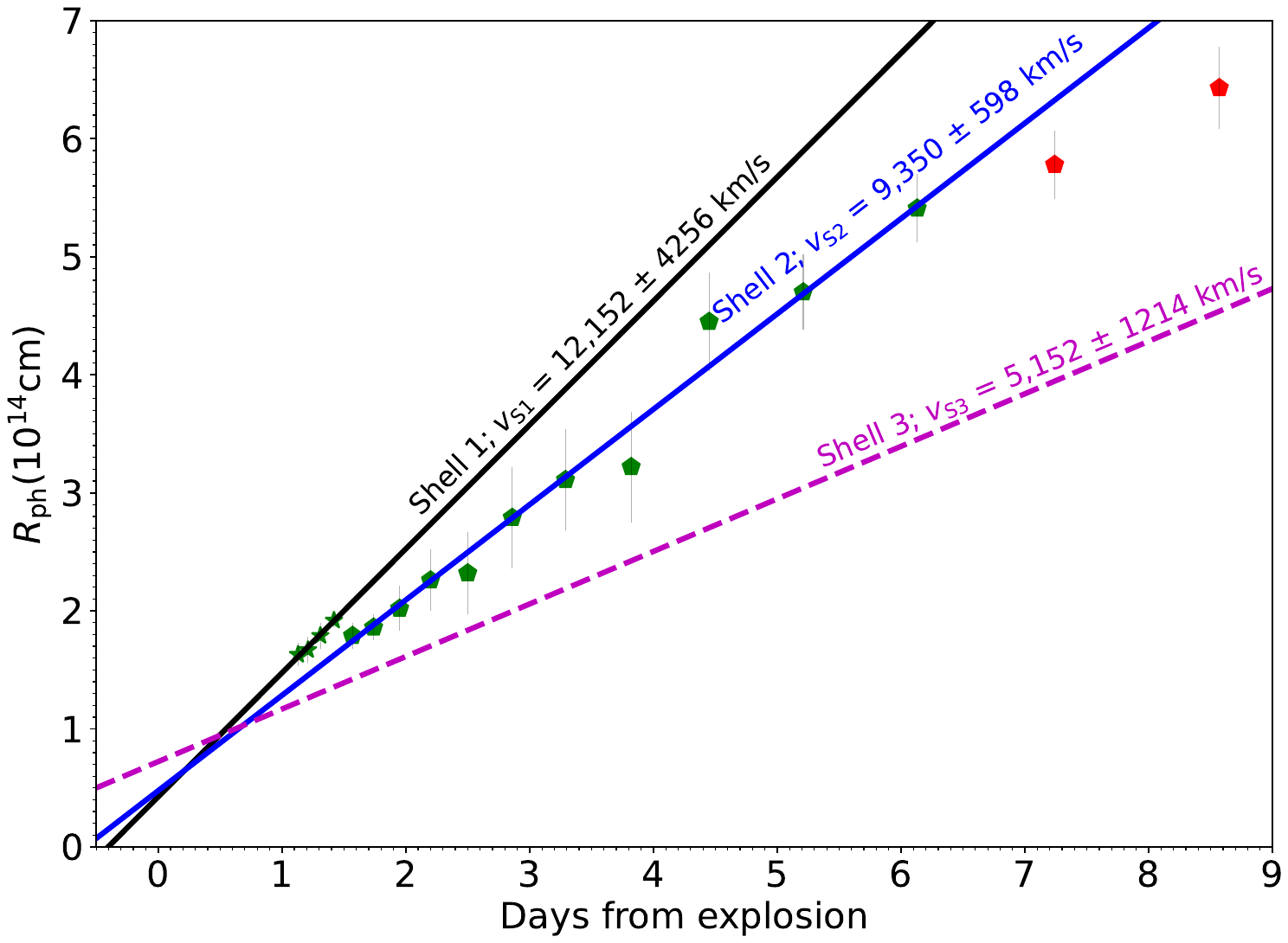}
\caption{The time evolution of the photospheric radius from \cite{Zimmermanetal2024} based on photometry (not on spectroscopy). We fit two straight lines through two groups of points: the first four points, green stars, at $t= 1.13-1.42 \dday$ that we mark as Shell 1, and the next 11 points, green pentagons, at $t= 1.57-6.13 \dday$ that we mark as Shell 2. The red pentagons are excluded from the fitting; they belong to the phase when the photosphere is moving inward in the mass coordinate of the ejecta. We also indicate the velocity of each shell during the fitting period. The line for Shell 3 fits a later time period, as shown in Figure \ref{fig:Graph2}. We identify these shells as photospheric shells appearing one after the other. The straight lines for the three shells are $R_{\rm ph1}= 1.050t +0.425$, $R_{\rm ph2}= 0.808t +0.477$, and $R_{\rm ph3}= 0.445t +0.723$, where the time is in days and the radius in units of $10^{14} \cm$. All lines have a coefficient of determination (R-Squared) of $R^2=0.98$. The velocities and their uncertainties are from fitting the ejecta photosphere radii without including the radii of the shells at $t=0$. Including the radii at $t=0$ changes the velocities little but substantially reduces the uncertainties, indicating that the three shells are distinct, by these photospheric radii (caveat: see text on the possibility that Shells 1 and 2 are a single shell).  
}
\label{fig:Graph1}
\end{center}
\end{figure*}

\begin{figure*}
\begin{center}
\includegraphics[trim=0.0cm 0.50cm 0.0cm 2.0cm ,clip, scale=0.53]{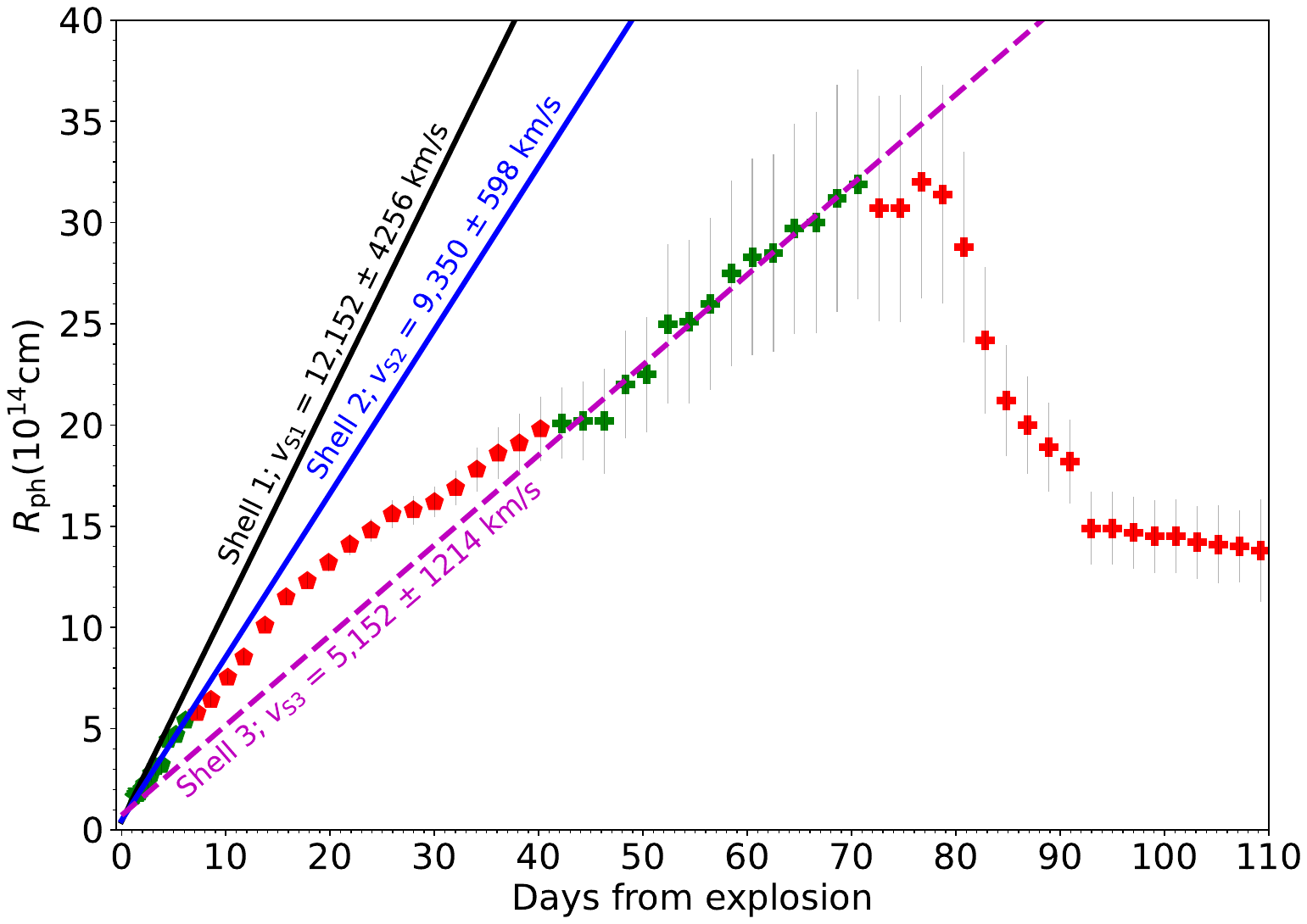}
\caption{Similar to Figure \ref{fig:Graph1}, but including later times, and different scales on the axes. As in Figure \ref{fig:Graph1}, red pentagons and pluses are excluded from fittings. We fit a third straight line to the points in the time period of  $t= 42.19-70.63 \dday$ (green pluses); this is Shell 3. 
 The points from day 7 to day 41 (red pentagons) are the photospheric radius in the ejecta inward to Shell 2 and outside Shell 3. During that time, the photosphere moves inward in the mass coordinate of the ejecta. Therefore, each point corresponds to a deeper mass layer than the previous point; the points do not represent a single layer. For that, we cannot fit a straight line through these points. Around day 45, the front of Shell 3 becomes the photosphere. From there to about day 71, a straight line corresponds to the front of Shell 3. After day 71, Shell 3 became transparent, and the photosphere moved inward in the mass coordinate of the ejecta (red pluses).  
}
\label{fig:Graph2}
\end{center}
\end{figure*}

We examine the photospheric radius $R_{\rm ph} (t)$ that \cite{Zimmermanetal2024} calculated in their thorough analysis of SN 2023ixf and give in a table: {\href{https://static-content.springer.com/esm/art%3A10.1038%2Fs41586-024-07116-6/MediaObjects/41586_2024_7116_MOESM3_ESM.xls}{Supplementary Data (Source Data Table 1)}}. 
Figure \ref{fig:Graph1} presents the photospheric radius that \cite{Zimmermanetal2024} calculated at early days (stars and pentagons) as in their Figure 2.
We use the same time scale as \cite{Zimmermanetal2024}, who took it from \cite{Yaronetal2023}, but note that the explosion time is uncertain.
The time $t=0$ corresponds to the first observed brightening due to the explosion. \cite{Yaronetal2023} estimated the uncertainty in the time to be about one hour. The explosion at the center of the star has started earlier. In Figure \ref{fig:Graph2}, we present the photospheric radii that \cite{Zimmermanetal2024} calculated at later times.  

\cite{Zimmermanetal2024} argued for a constant photosphere until $t \simeq 2.75 \dday$. We instead find that there are two photospheric shells during that time. We fit two straight lines to the points in Figure \ref{fig:Graph1}. Shell 1 that we fit through the first four point (green stars) corresponds to a velocity of $v_{\rm S1}=12,152 \pm 4256 \km \s^{-1}$, and Shell 2 that we fit through the green pentagons, to  $v_{\rm S2}=9,350 \pm 598 \km \s^{-1}$;   these lines are the fit through the points of the ejecta photosphere only, not including the photosphere (stellar radius) at $t=0$. 
In Figure \ref{fig:Graph2}, we extend the time, and find a time period, represented by the green pluses, where the photosphere increases linearly, this is Shell 3 with a velocity of $v_{\rm S3}=5,152 \pm 1214 \km \s^{-1}$  as fitted through the points of photospheric radii but not through the center. The three fitted lines through the ejecta photospheric radii $R_{\rm ph} (t)$ cross the expected stellar radius at $t=0$, as should be for expanding ballistic (more or less) shells.    

Shell 1 expands, and the photosphere moves outward with it. As it starts to become transparent (optically thin), the photosphere expands more slowly as it recedes in the ejecta's mass coordinate. In the transition from Shell 1 to Shell 2, the photosphere also decreases in radius, suggesting that Shell 1 carries little mass. Then the slower, denser Shell 2 catches up with the photosphere and becomes the photosphere, with linear growth. Later, Shell 2 becomes optically thin, and the photosphere expansion slows down again as it recedes in the ejecta. At about $t \simeq 45 \dday$, Shell 3 catches up and makes the photosphere. The photosphere expands linearly until Shell 3 becomes optically thin at $t \simeq 71 \dday$. 

We make the following comments: 
 (1) When we draw the first two shells just by the photospheric points, 4 points for Shell 1 and 11 points for Shell 2, the velocities have large uncertainties $v_{\rm S1}=12,152 \pm 4256 \km \s^{-1}$ and $v_{\rm S2}=9,350 \pm 598 \km \s^{-1}$. 
However, this fit is intended to show that the straight lines pass through the center of the explosion. To derive the shell velocity with the correct uncertainties, we must include the explosion's center. We do not know exactly the stellar radius or the time the shells broke from the photosphere. We set the same initial radius for the three shells at $t=0$, rather than a different initial radius for each shell, although it is clear they did not all pass through that radius at the same time. For a shells' radius of $R_{6}(0) = 650 R_\odot \pm 200 R_\odot$ at $t=0$ the velocities change little, but the uncertainties are much smaller now: $v_{\rm S1r6}=11,911 \pm 1,231 \km \s^{-1}$, $v_{\rm S2r6}=9,391 \pm 477 \km \s^{-1}$, and $v_{\rm S3r6}=5213 \pm 193 \km \s^{-1}$;  for $R_{7}(0) = 750 R_\odot \pm 250 R_\odot$ the velocities are $v_{\rm S1r7}=11,381 \pm 1483 \km \s^{-1}$, $v_{\rm S2r7}=9,280 \pm 507 \km \s^{-1}$, and $v_{\rm S3r7}=5,197 \pm 195 \km \s^{-1}$; 
Subscripts $r6$ and $r7$ denote the two shells' radii at $t=0$ that we use, respectively. 
For larger initial shells' radii than $\simeq 750 R_\odot$ (at $t=0$), the velocities of Shells 1 and 2 depart more from the velocities of the points alone (the velocities given in Figures \ref{fig:Graph1} and \ref{fig:Graph2}), and so we regard these shells' radii at $t=0$ as less likely. 
Shells 1 and 2 are about one sigma apart, and Shells 2 and 3 are about six sigma apart. These velocities, along with their uncertainties, reinforce the hypothesis that the system comprises of at least two shells (1+2 as one shell, and Shell 3), and with some uncertainty from three shells. 

(2) From Figure \ref{fig:Graph2}, one might propose that there are two shells, where Shell 1 and Shell 2 are part of the same shell. Shell 3 is the same as before. We find that a straight line fits the first 22 points from $t=1.13 \dday$ to $t=17.81 \dday$. Namely, Shell 1, Shell 2, and the following 7 points form one photospheric shell. The velocity of such a massive early shell is $v_{\rm SE}=7,676 \pm 130 \km \s^{-1}$.  This velocity is very similar to the $8,000 \km \s^{-1}$ early-time ejecta velocity that \cite{Zimmermanetal2024} deduce. The velocity similarity supports a massive shell. In addition, the first four points that make Shell 1 do not have UV observations, and therefore, the fitting of the black-body temperature and radius is less certain. Given the photospheric radii of Shell 1 as calculated by \cite{Zimmermanetal2024}, we consider the interpretation of Shell 1 and 2 as separate shells more likely. However, if the temperatures of these four points are underestimated and their photospheric radii are smaller, causing them to lie along the line of Shell 2, then Shell 1 and 2 would merge into a single Shell. In that case, the interpretation of \cite{Zimmermanetal2024} of a constant photosphere until $t \simeq 2.75 \dday$ does not hold. Overall, we caution that Shell 1 and Shell 2 may not be separate shells, but rather a single Shell.    
Even under this interpretation, in which Shells 1 and 2 are one massive shell, Shell 3 is separate, and the shells model exists with two shells. 

(3) A shell in this study might be a complete shell covering a solid angle of $\Omega_{\rm sh} = 4 \pi$ around the explosion site, or a segment of a shell, a cap (or pairs of caps), covering much less, down to $\Omega_{\rm sh} < 2 \pi$. The cap should be along the line of sight and be large enough to cover a large fraction of the supernova shell towards an observer along its direction. Crudely, a cap should cover an angle of $\gtrsim 0.5 \pi$, for which it covers a half of a half sphere to an observer along its axis, to form such a photospheric shell. 

(4) \cite{Tylenda2005} found three shells for the luminous red nova V838~Mon. In the case of V838~Mon, the photosphere decreases in radius between the shells. In our analysis, we find a decrease between Shell 1 and Shell 2. This decrease is covered by two observational points (transition from green stars to green pentagons in Figure \ref{fig:Graph1}). In the transition from Shell 2 to Shell 3, the photosphere expansion slows down, but its radius does not decrease; the transition occurs over a relatively long time, $t \simeq 6 \dday$ to $t \simeq 45 \dday$. Namely, the transition from Shell 2 to Shell 3 is gradual, and the photosphere moves inward only slightly in mass coordinate before Shell 3 catches up.  This gradual and small deepening of the photosphere in mass coordinate explains the small change in the slope of $R_{\rm ph} (t)$ between Shell 2 and Shell 3 (Figure \ref{fig:Graph2}), and the continuous slope in SN 2023ixf luminosity in that phase. 

(5) At late times, a free shell of ejecta expands at a constant velocity (if it does not collide with a massive circumstellar matter), as its kinetic energy is much larger than its thermal energy. This is not true very close to the origin, when the ejecta is hot and thermal energy is convected to kinetic energy (adiabatic cooling). Since the stellar radius is $\approx 10^{14} \cm$, the velocity of the shells close to the center, $r \lesssim 2 \times 10^{14}$, is not constant. A collision with the dense, compact circumstellar matter might slow the first shells, while the conversion of thermal to kinetic energy accelerates them near the center. The study of the shell's behavior close to the star is a subject for a separate study. 

\section{CCSN remnants with shells}
\label{sec:Remnants}

Some CCSNRs present two or three shells, where the shells can be complete (or almost complete), covering a solid angle of $\Omega_{\rm sh} = 4 \pi$, or be partial, covering a solid angle of $\Omega_{\rm sh} < 2 \pi$. Such shells, if formed during the explosion, can lead to the formation of two or more photospheric shells during the photospheric phase of the explosion.  We present four CCSNRs, all attributed to the JJEM, that show such shell morphologies. They suggest that energetic jet pairs can form such shells, as the JJEM predicts in some cases (but not in all).  

\begin{figure}[th]
\begin{center}
\includegraphics[trim=0.0cm 10.05cm 6.8cm 2.2cm ,clip, scale=0.83]{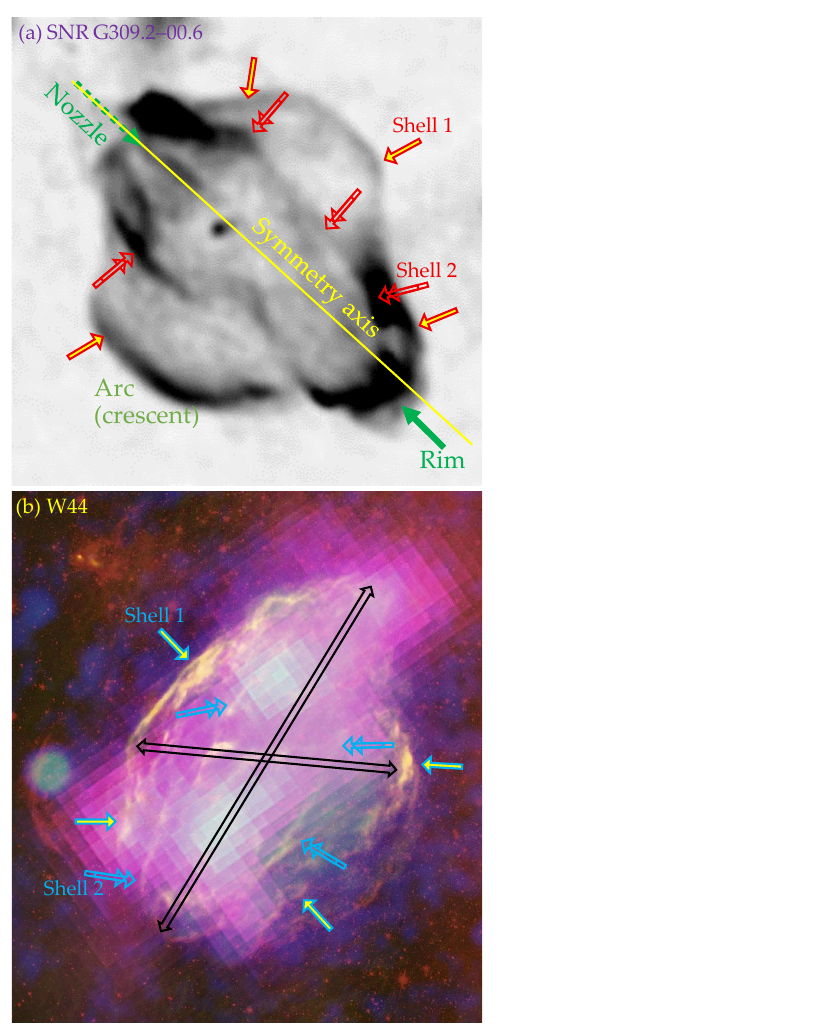}
\caption{ Two images of CCSN remnants with two full (or almost full) shells (but not spherical). Studies attributed their morphologies to the JJEM. In each image, we marked the two prominent shells that can form two photospheric shells one after the other during the photospheric phase of such CCSNe; the arrows point to only four points on the projection of each shell boundary onto the plane of the sky. Filled double-lined arrows point at the first shell, and the empty double-lined double-headed arrows point at the second shell. (a) A radio image of SNR G309.2–00.6 adapted from \cite{Gaensleretal1998}, who already argued that jets shaped this CCSN. The rim-nozzle symmetry marks are from \cite{Soker2024PNSN}. (b) A composite image of CCSNR W44 from 
\href{https://www.nasa.gov/universe/nasas-fermi-proves-supernova-remnants-produce-cosmic-rays/}{a NASA site}. \cite{Soker2024W44} added the two double-headed arrows that depict the two axes of the two energetic pairs of jets that participated in the explosion of W44. 
Magenta: GeV gamma-ray from Fermi’s LAT; yellow: Radio from the Karl G. Jansky Very Large Array; Red: infrared; Blue: X-ray from ROSAT. 
Credit: NASA/DOE/Fermi LAT Collaboration, NRAO/AUI, JPL-Caltech, ROSAT. 
}
\label{fig:MorphologiesFull}
\end{center}
\end{figure}

In Figure \ref{fig:MorphologiesFull} we present two CCSNRs, each with almost full two shells. We mark four points at the boundary of each shell as projected on the plane of the sky with arrows: Shell 1 is the first to form a photosphere, and then Shell 2 catches up with the photosphere of the first shell, and takes over with a slower expansion velocity.  
Both CCSNRs show clear signatures of at least two pairs of jets. 
Already \cite{Gaensleretal1998} argued that SNR G309.2-00.6 was shaped by jets (panel a of Figure \ref{fig:MorphologiesFull}). Based on its morphology, \cite{Soker2024PNSN} and \cite{ShishkinKayeSoker2024} attributed the explosion of SNR G309.2-00.6 to the JJEM. \cite{Soker2024W44} argued that two energetic pairs of jets left their clear marks on CCSNR W44 (panel b of Figure \ref{fig:MorphologiesFull}). According to the JJEM, more pairs of jets, but with lower energies, participated in the explosion process of W44. The two pairs of jets that \cite{Soker2024W44} mark (two double-sided arrows) inflated the two shells that we mark here. This, along with the other images, suggests that two, and possibly three, energetic pairs of jets contributed to the explosion of SN 2023ixf. The JJEM predicts such structures as three-dimensional hydrodynamical simulations show (e.g., \citealt{Braudoetal2025}). 

\begin{figure}[th]
\begin{center}
\includegraphics[trim=0.0cm 9.3cm 7.8cm 0.0cm ,clip, scale=0.64]{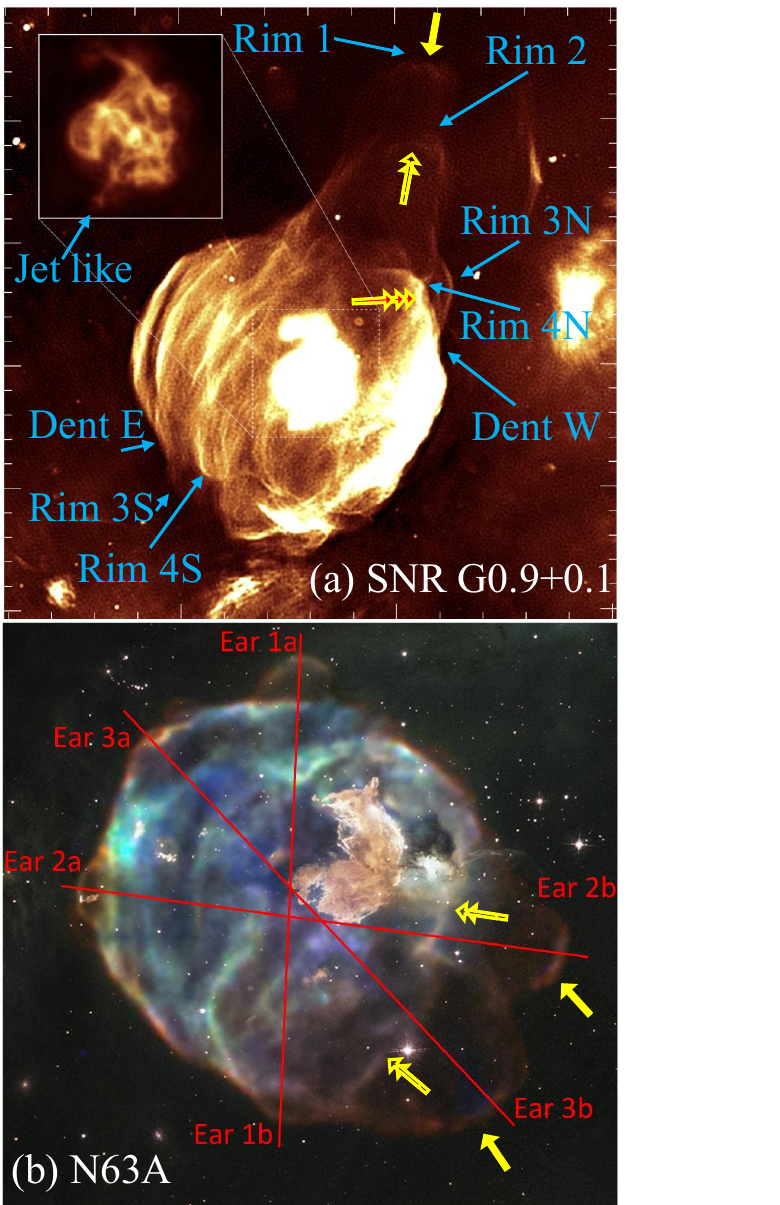}
\caption{ Two images of CCSNRs where at least one shell is partial. 
(a) A MeerKat radio image at 1.28 GHz of SNR~G0.9+0.1 adapted from \cite{MeerKAT2022}. The inset on the upper left is a desaturated image of the pulsar wind nebula. \cite{Soker2025G0901} added the pale-blue marks of structural features used to identify the point-symmetric morphology. For a line of sight along the large ear in the north, the three shells (their projection on the plane of the sky forms the rims) might form three photospheric shells. The first two (pointed at by the solid one-head yellow arrow and double-lined double-headed yellow arrow) are partial. The third shell (pointed at by the three-headed arrow) is the main shell of the SNR.  
(b) A Chandra X-ray image of N63A (red, green, blue for different X-ray energy bands); see also \cite{Karagozetal2023}. 
\cite{Soker2024CounterJet} added the three red lines between the tips of opposite ears to mark the symmetry axis of three pairs of jets that participated in the explosion of this point-symmetric CCSNR. Each ear's front can form the first photospheric shell, depending on the viewing angle. The main CCSNR forms the next photospheric radius.   
The light brown region to the upper right of the three red lines is optical light detected by Hubble. (Credit: Enhanced Image by Judy Schmidt based on images provided courtesy of NASA/CXC/SAO \& NASA/STScI.)
}
\label{fig:MorphologiesPartial}
\end{center}
\end{figure}

In Figure \ref{fig:MorphologiesPartial} we present two CCSNRs with partial shells. Namely, they can lead to the observations of two or more photospheric shells only from specific viewing angles. In panel (a), we present SNR~G0.9+0.1. \cite{Soker2025G0901} identified a point symmetric morphology by the structural features marked on the image in pale-blue. We refer to the large ear to the north that has two large front rims. Rims 1 and 2 are the projections of caps, or partial shells, on the plane of the sky. Such rims are formed by jets. If an observer is located along and near the ear direction, the caps of Rim 1 (pointed at by the one-headed yellow arrow) and Rim 2 (double-headed arrow) will form two photospheric shells. The main shell of this CCSNR (pointed at by the triple-headed arrow) will form a third photosphere.
 In panel (b), we present an image of the CCSNR N63A, with its three pairs of ears that support the JJEM \citep{Soker2024CounterJet}. Ears 2b and 3b are large enough to form large shells at their front (pointed at with the solid yellow arrows). The fronts of these two ears might form a first photospheric shell for an observer facing them. The main shell of N63A (pointed at with the two double-headed arrows) will form the second photospheric shell.    

In addition to these four CCSNe, the Vela CCSNR has two full shells, and SNR~G107.7-5.1 has an ear with three rims; the morphologies of both CCSNRs are attributed to the JJEM  
(Vela: \citealt{Soker2023SNRclass, SokerShishkin2025Vela}; G107.7-5.1: \citealt{Soker2024CF}). About twenty CCSNR morphologies were directly attributed to the JJEM (most others have two dispersed morphologies to be classified). We find six with prominent multiple shell structure.   
We conclude that jet-shaped morphologies can form two or three, and possibly more, shells during the explosion and lead to the appearance of multiple photospheric shells. 

\section{Summary} 
\label{sec:Summary}

Examining the photospheric time evolution of SN 2023ixf that \cite{Zimmermanetal2024} calculated, we identify three evolutionary time periods with a constant expansion velocity, as we mark in Figures \ref{fig:Graph1} and \ref{fig:Graph2}.  However, the first two segments, Shells 1 and 2, might be a single shell.  We attribute these to shells of the SN 2023ixf ejecta. Our calculated velocities of these shells during their respective photospheric phases are $v_{\rm S1} \simeq 12,150 \km \s^{-1}$, $v_{\rm S2} \simeq 9,350 \km \s^{-1}$, and $v_{\rm S3}\simeq 5,150 \km \s^{-1}$.
The velocity of a single shell that combines Shells 1 and 2 and up to day 17, is $v_{\rm SE}=7,676 \pm 130 \km \s^{-1}$.   The shells can be complete (covering a solid angle of $\Omega_{\rm sh} = 4 \pi$), or partial if the partial shell, or cap, is along the line of sight.  

Several CCSNRs have morphologies with two or more complete or partial shells. We presented four of these in Figures \ref{fig:MorphologiesFull} and \ref{fig:MorphologiesPartial}, and mention two more in Section \ref{sec:Remnants}. Earlier studies attributed all these CCSNRs to the JJEM, primarily to two or three energetic jet pairs. Additionally, three-dimensional hydrodynamic simulations \citep{Braudoetal2025} demonstrate the formation of such structures by consecutive pairs of jets. 

We caution that there are uncertainties in the values of the photospheric radii that \cite{Zimmermanetal2024} calculated. The four points of Shell 1 lack UV observations, and Shells 1 and 2 may be a single, more massive shell. In that case, SN 2023ixf contains two shells rather than 3. The uncertainties of the points of Shell 3 are large. It is possible to draw a line through these uncertainties that are not straight. In that interpretation, there is no Shell 3. 
We limit this study to interpreting the implications of the photospheric radii as calculated by \cite{Zimmermanetal2024}. \textit{ In any case, one of the main points of our study is the claim that the structure of shells that many CCSNRs reveal, might be detected in the photospheric phase of CCSNE. }  

Under the interpretation of three shells that we adopt here,  Shell 1 of SN 2023ixf becomes optically thin (transparent) very early, at $t=1.4 \dday$, and the photospheric radius decreases (Figure \ref{fig:Graph1}). This shell, therefore, contains little mass. This shell may be the front of the ejecta, which has little mass and is not necessarily jet-shaped. However, Shell 2 and Shell 3 are more massive, and, based on CCSNRs that we present in Section \ref{sec:Remnants}, we attribute their existence to energetic jets.  

Based on the morphologies of CCSNRs, future modeling of shells must consider non-spherical shells, which we did not address in this study. The modeling should also address polarization, as observation finds in SN 2023ixf \citep{Vasylyevetal2025}; polarization modeling must also consider non-spherical circumstellar matter.    

We state that the structure of the two, or possibly three,  photospheric shells that we identified in SN 2023ixf is fully compatible with jet-shaped CCSNRs and simulations. We take this to support the JJEM. This study adds to the accumulating evidence that the JJEM is the primary explosion mechanism of CCSNE. 

\section*{Acknowledgements}

  We thank an anonymous referee for very useful comments. 
NS thanks the Charles Wolfson Academic Chair at the Technion for the support.


%
\bibliography{reference}{}
\bibliographystyle{aasjournal}
  


\end{document}